\documentclass[aps,pra,superscriptaddress,amsmath,amssymb,
twocolumn
]
{revtex4-1}

\usepackage{
graphicx,
empheq,
hyperref,
placeins, 
xcolor}
\usepackage[capitalise]{cleveref}

\graphicspath{{./images/}{./figures/}}

\begin{document}
\title{Graphene plasmon-phonon coupled modes at the exceptional point}
\author{Sang Hyun Park}
\email{park2451@umn.edu}
\affiliation{Department of Electrical \& Computer Engineering, University of Minnesota, Minneapolis, Minnesota, 55455, USA}
\author{Shengxuan Xia}
\affiliation{Department of Electrical \& Computer Engineering, University of Minnesota, Minneapolis, Minnesota, 55455, USA}
\affiliation{School of Physics and Electronics, Hunan University, Changsha, 410082, China}
\author{Sang-Hyun Oh}
\affiliation{Department of Electrical \& Computer Engineering, University of Minnesota, Minneapolis, Minnesota, 55455, USA}
\author{Phaedon Avouris}
\affiliation{Department of Electrical \& Computer Engineering, University of Minnesota, Minneapolis, Minnesota, 55455, USA}
\affiliation{IBM Thomas J. Watson Research Center, Yorktown Heights, New York, 10598, USA}
\author{Tony Low}
\email{tlow@umn.edu}
\affiliation{Department of Electrical \& Computer Engineering, University of Minnesota, Minneapolis, Minnesota, 55455, USA}

\begin{abstract}
	{Properties of graphene plasmons are greatly affected by their coupling to phonons. While such coupling has been routinely observed in both near-field and far-field graphene spectroscopy, the interplay between coupling strength and mode losses, and its exceptional point physics has not been discussed. By applying a non-Hermitian framework, we identify the transition point between strong and weak coupling as the exceptional point. Enhanced sensitivity to perturbations near the exceptional point is observed by varying the coupling strength and through gate modulation of the graphene Fermi level. Finally, we also show that the transition from strong to weak coupling is observable by changing the incident angle of radiation.}
\end{abstract}

\maketitle
\section{Introduction}
%Survey of coupling behavior in physical systems
\par The response of coupled systems have been shown to strongly depend on the relative strength of their coupling and losses. Here we will be focusing on light-matter coupling. When the coupling is strong relative to the losses, the mode degeneracy is lifted, giving rise to hybridized modes at split frequencies\cite{Liu2014}. The resulting absorption spectrum has two peaks that are separated by $\Delta\omega\gg\gamma$, where $\gamma$ is the linewidth. When the coupling is weak, the peak's degeneracy is no longer lifted but these degenerate modes can still interfere. In the limit of contrasting linewidths ($\gamma_{1}\gg\gamma_{2}$), destructive interference between the modes results in a sharp optical transparency region within a broad absorption peak\cite{Fleischhauer2005}. Such `strong' and `weak' coupling response has been observed in many physical systems including three-level atoms\cite{Anisimov2011}, whispering-gallery-mode microresonators\cite{Peng2014a}, mechanical systems\cite{Tassin2013}, and plasmonics\cite{Zhang2008a, Tormo2015a}.

%Parity-time symmetric systems
\par The role of coupling($\kappa$) and loss($\gamma$) is also an important factor in non-Hermitian sytems\cite{El-Ganainy2018}, which are open systems that exchange energy with their environment. The importance of coupling and loss is highlighted in a subset of non-Hermitian systems that obey parity-time (PT) symmetry\cite{Guo}. When $\kappa/\gamma>1$, the eigenstates of the system are PT symmetric and the eigenvalues are real. When $\kappa/\gamma<1$ the eigenstates are no longer PT symmetric and the eigenvalues become complex. At $\kappa=\gamma$, there is a non-Hermitian singularity, also known as the exceptional point (EP), where both the eigenvalues and eigenvectors coalesce (see \cref{fig:eigensurface}). This transition has been utilized to demonstrate non-reciprocal propagation in waveguides\cite{Ruter2010}, asymmetric transmission of polarized light\cite{Lawrence2014}, and PT-symmetric lasers\cite{Hodaei2014}. Also, exceptional points have been reported to exhibit enhanced sensitivity to perturbations\cite{Wiersig2020a,Hodaei2017, Chen2017}.

% Plasmon phonon coupling in graphene
\par Plasmons in graphene couple with phonons and have been experimentally observed in various contexts. When patterned graphene is placed on a silicon dioxide substrate, coupling with the surface optical phonon modes are directly observable in the extinction spectra\cite{Yan2013}. Coupling with surface modes in atomically thin polar materials such as hBN was also observed\cite{Brar2014}. For bilayer graphene, plasmons couple with its intrinsic $\Gamma$ point optical phonon, resulting in a sharp phonon-induced transparency\cite{Yan2014, Bezares2017}. Finally, the high confinement of graphene plasmons also allows them to couple with vibrational modes of adsorbed molecular layers or gases, allowing their fingerprinting through its extinction spectrum\cite{Rodrigo2015, Farmer2016, Hu2019, Khaliji2020, Lee2019}.

In this paper, we discuss graphene plasmon-phonon coupling from the perspective of a non-Hermitian system, highlighting the interplay between coupling and loss. Harmonic inversion analysis\cite{Fuchs2014} of the absorption spectra allows us to precisely identify the non-Hermitian singularity called the exceptional point positioned within the parameter space. By quantifying  the changes in the spectral features upon perturbation of coupling strength, incidence angle of light, dielectric environment, and the position of the Fermi level, we demonstrate that the spectral splitting due to plasmon-phonon coupling in graphene is indeed most sensitive to perturbation when the system is at the exceptional point. Non-Hermitian and exceptional point physics provides a unique perspective in the understanding of plasmon-phonon coupled systems, and the approach we outline here sets the stage for the future design of sensing plasmon-phonon coupling characteristics via the EP.

\section{Exceptional point physics in plasmon-phonon coupling}
\begin{figure}
	\centering
	\includegraphics[width=0.5\textwidth, height=0.75\textwidth]{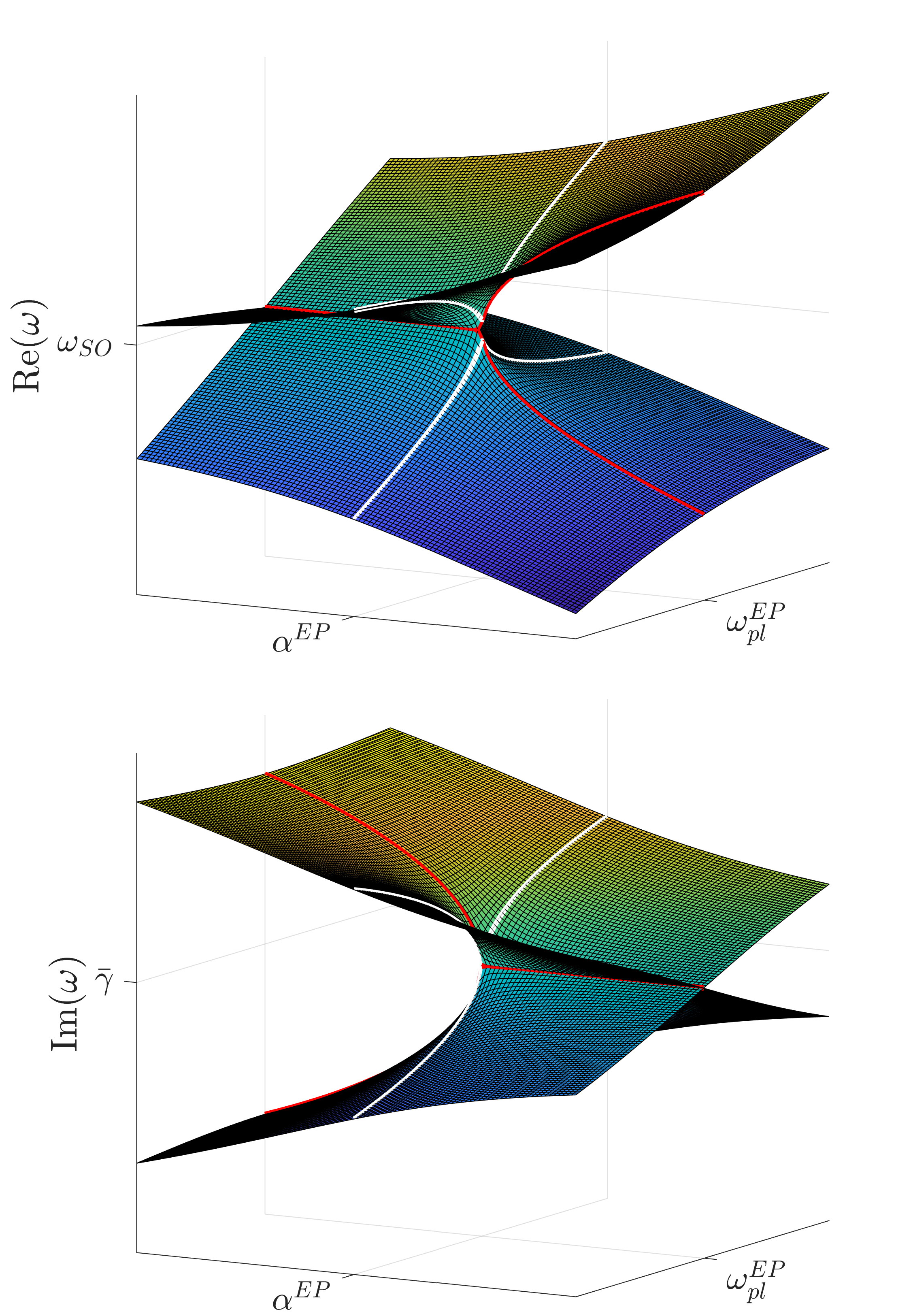}
	\caption{Eigenvalue surfaces calculated from \cref{eq:rpa_dielectric} in the $(\omega_{pl},\alpha)$ parameter space. The red(white) line shows a path along the surface with $\omega_{pl}=\omega_{pl}^{EP}$($\alpha=\alpha^{EP}$).}
	\label{fig:eigensurface}
\end{figure}
Exceptional point physics in plasmon-phonon coupled system can be elucidated from both a classical and quantum point-of-view. As an example, we consider graphene on a polar substrate with surface optical phonon mode $\omega_{SO}$\cite{Hwang2010}
% Classical coupled mode model and relation to PT symmetry
\par We begin with a simple coupled mode model describing the response of the plasmon-phonon coupled system\cite{Haus1985}. The equations of motion for the plasmon($a(t)$) and phonon($b(t)$) amplitudes are
\begin{align} \label{eq:coupled_mode_equations}
  \dot{a}(t) &= i\omega_{pl}a-\gamma_{pl}a + i\kappa b + \sqrt{\gamma_{pl,e}}s_{a,\text{in}} \\
  \dot{b}(t) &= i\omega_{SO}b-\gamma_{SO}b + i\kappa a
\end{align}
where $\gamma_{pl} = \gamma_{pl,e} + \gamma_{pl,i}$ accounts for plasmon loss including intrinsic($\gamma_{pl,i}$) and extrinsic ($\gamma_{pl,e}$) contributions, $\gamma_{SO}$ is the phonon loss, $\omega_{pl}$ and $\omega_{SO}$ are the resonant frequencies of the plasmon and phonon, $\kappa$ is the near-field coupling strength, and $s_{a,\text{in}}$ quantifies the input field that couples into the plasmon. Here, the surface optical(SO) phonon modes are not directly excited by light since they are not infrared active. Assuming that the amplitudes have a $e^{i\omega t}$ time dependence, the response of the plasmon is given by
\begin{equation}\label{eq:plasmon_response}
  a(t) = \frac{i\sqrt{\gamma_{pl,e}}s_{a,\text{in}}(\omega-\tilde{\omega_{SO}})}{\kappa^2 - (\omega-\tilde{\omega}_{pl})(\omega-\tilde{\omega}_{SO})}e^{i\omega t}
\end{equation}
where we have defined $\tilde{\omega}_{pl/SO}=\omega_{pl/SO}+i\gamma_{pl/SO}$. From \cref{eq:plasmon_response} a resonant response occurs when $\omega_{\pm}=(\tilde{\omega}_{pl}+\tilde{\omega}_{SO})/2 \pm \sqrt{(\tilde{\omega}_{pl}-\tilde{\omega}_{SO})^{2}+4\kappa^{2}}/2$. When $\omega_{pl}=\omega_{SO}=\omega_{0}$, the poles are at $\omega_{\pm}=\omega_{0}+i\bar{\gamma} \pm \sqrt{\kappa^{2}-\gamma^{2}}$, where we have defined $\bar{\gamma}=(\gamma_{pl}+\gamma_{SO})/2$ and $\gamma = (\gamma_{pl}-\gamma_{SO})/2$. Written in this form, we see that there is a transition at the poles for $\kappa=\gamma$. The poles found from \cref{eq:plasmon_response} can be identified with the eigenvalues of a PT-symmetric matrix\cite{Bender1998} of the form
\begin{equation}
	\hat{H}_{PT}=
	\begin{pmatrix}
		\omega_{pl} + i\gamma_{pl} & \kappa \\ \kappa & \omega_{SO} + i\gamma_{SO}
	\end{pmatrix}.
\end{equation}
By making this connection, we see that the transition point at $\kappa=\gamma$ could be identified as the exceptional point. If we consider a system in which the loss is fixed and coupling is tunable, writing $\kappa=\gamma + \Delta\kappa$ gives the eigenvalues
\begin{equation}\label{eq:eigenvalues}
    \omega_{\pm}=\omega_0+i\bar{\gamma} \pm \sqrt{\Delta\kappa}\sqrt{\Delta\kappa + 2\gamma}.
\end{equation}
This shows the well known characteristic square-root dependence of the mode splitting due to perturbations near an EP\cite{Wiersig2020a}. 

% QM description of the coupled system and relation to PT symmetry
\par Next, we explore the relationship between plasmon-phonon coupling and the EP via a quantum approach. In the limit $\hbar\omega\ll E_F$, the total dielectric function for graphene within the random phase approximation including electron coupling to the SO phonons can be described by\cite{Hwang2010}
\begin{equation}\label{eq:rpa_dielectric}
  \epsilon(q, \omega) = 
    1 - \frac{\omega^{2}_{pl}(q)}{(\omega + i\gamma_{gr})^{2}} - 
    \frac{\alpha\omega_{SO}^{2}}{(\omega+i\gamma_{SO})^{2}-\omega_{SO}^{2}+\alpha\omega_{SO}^{2}}
\end{equation}
where $\omega_{pl}^{2}(q)=q|E_{F}|e^{2}/2\pi\hbar^{2}\epsilon_{0}$ is the plasmon frequency, $\gamma_{gr}$ is the electron scattering rate of graphene, and $\alpha = e^{-2qd}[(\epsilon_{\infty}+1)^{-1}-(\epsilon_{st}+1)^{-1}]$ quantifies the strength of the electron-phonon interaction (analogous to $\kappa$ defined in the coupled mode model). The dispersion of the coupled modes can be found by solving $\epsilon(q, \omega)=0$. Solutions of \cref{eq:rpa_dielectric} as a function of $\alpha$ and $\omega_{pl}$ are shown in \cref{fig:eigensurface}. The topology of the solutions reveal a point in the parameter space in which the real and imaginary parts of the solution become simultaneously degenerate, thus demonstrating the existence of an EP. Moreover, perturbations to the system at the EP induce splitting of the eigenvalues proportional to the square-root of the perturbation strength\cite{Wiersig2020a}.
\begin{figure}
	\centering
	\includegraphics[width=0.5\textwidth, height=13cm]{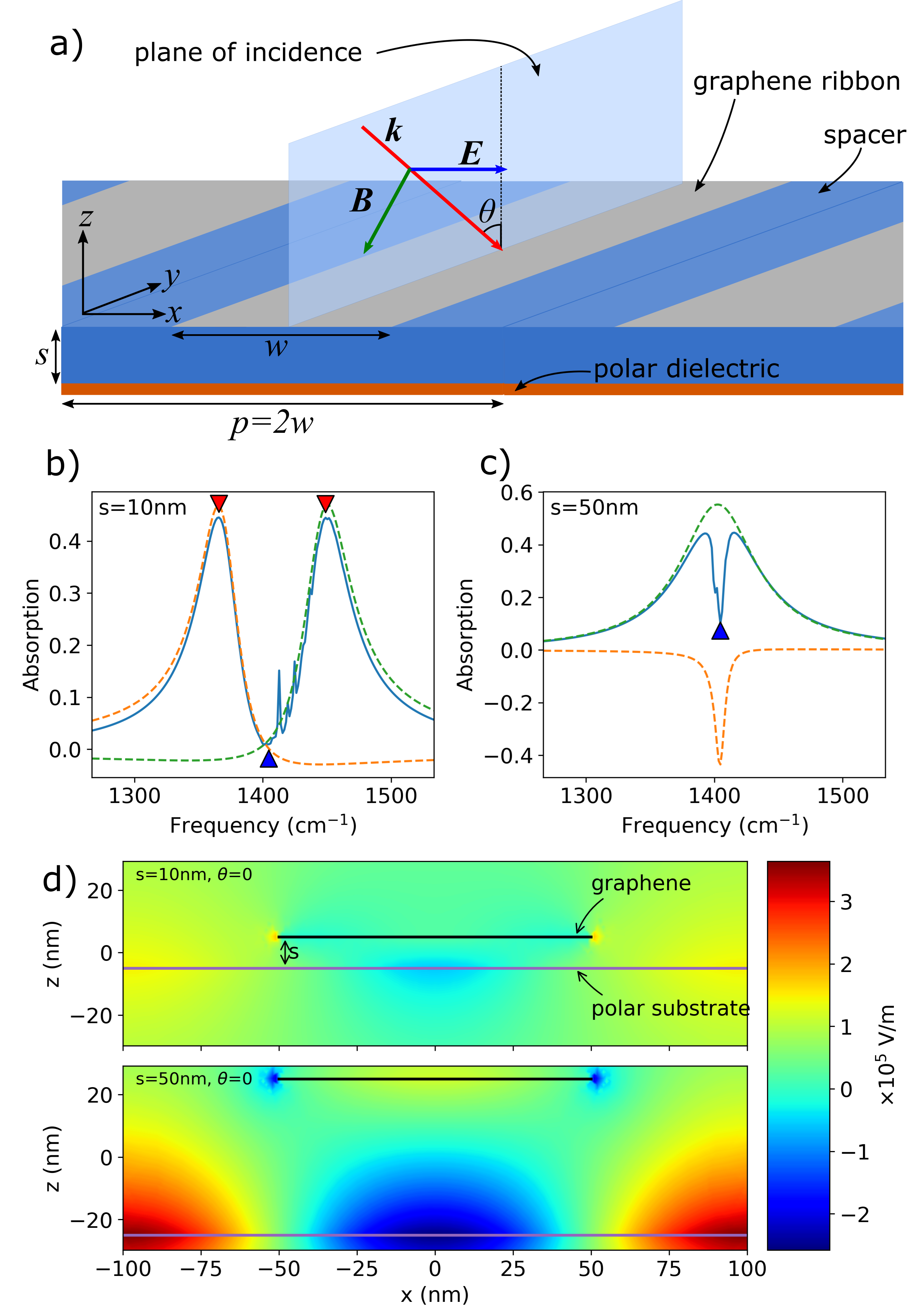}
 	\caption{(a) Schematic of the graphene ribbon/polar substrate structure. By patterning graphene into ribbons of width $w=100$nm and periodicity $p=200$nm we excite plasmons with $q=\pi/w$\cite{Nikitin2014} under incident light polarized along the width of the nanoribbons. Absorption spectra for (b) strong coupling($s=10$nm) and (c) weak coupling($s=50$nm). The dashed lines show results of decomposition using the harmonic inversion analysis method. Finer peaks near 1400cm$^{-1}$ are from higher order modes between the plasmon and phonon. (d)$E_x$ fields for strong (s=10nm) and weak (s=50nm) coupling are shown, both at frequency $\omega=\omega_{a}=\omega_{b}$ and $\theta=0$. }
 	\label{fig:schematic}
\end{figure}

\section{Mode splitting at exceptional point}
% Graphene modeling
\par Plasmon-phonon coupling in graphene can be studied through a far-field scattering experiment via its absorption spectrum. A typical device consisting of an array of graphene nanoribbons as illustrated in \cref{fig:schematic}(a). Graphene was modeled as a surface current density with conductivity given by
\begin{equation}\label{eq:drude_conductivity}
  \sigma_{\text{gr}}(\omega) = \frac{\sigma_{0}}{\pi}\frac{4E_{F}}{\hbar\gamma_{gr}-i\hbar\omega}
\end{equation}
where $\sigma_{0} = e^{2}/4\hbar$. The lifetime of carriers in graphene is set to $\tau_{gr}=\gamma_{gr}^{-1}=100$fs\cite{Yan2013}. The polar dielectric material is modeled using the dipole oscillator model
\begin{equation}
  \epsilon_{\text{sub}}(\omega)=\epsilon_{\infty}+(\epsilon_{st}-\epsilon_{\infty})\frac{\omega_{TO}^{2}}{\omega_{TO}^{2}-\omega^{2}-i\gamma_{TO}\omega}.
\end{equation}
where $\omega_{TO}$ is the transverse optical phonon frequency, $\gamma_{TO}$ is the phonon scattering rate, and $\epsilon_{st}(\epsilon_{\infty})$ are the low(high) frequency limits of the dielectric function. A polar dielectric of thickness $d$ can be shown to support surface optical phonon modes with the thickness dependent dispersion\cite{Goncalves2016}
\begin{equation}\label{eq:surface_dispersion}
	\omega_{+}\ :\ \tanh(\mu_{2}d/2) = -\frac{\epsilon_{2}\mu_{1}}{\epsilon_{1}\mu_{2}},\quad
	\omega_{-}\ :\ \coth(\mu_{2}d/2) = -\frac{\epsilon_{2}\mu_{1}}{\epsilon_{1}\mu_{2}}
\end{equation}
where $\epsilon_{2}=\epsilon_{sub}(\omega)$, $\epsilon_{1}$ is the dielectric constant of the surrounding medium, and $\mu_{1,2}=\sqrt{\epsilon_{1,2}\omega^2/c^2-q^2}$. In the calculations and simulations to follow, we set $\omega_{TO}=1400$cm$^{-1}$, $\tau_{SO}=\gamma_{SO}^{-1}=5$ps, $\epsilon_{\infty}=6.4$, and $\epsilon_{st}=4.95$. For a thin polar dielectric, the SO phonon mode that the plasmon couples to will be very close to the TO phonon frequency. A more detailed discussion of the surface optical phonon mode dispersion is given in the Supplemental Information (SI). Here, we assume that the polar dielectric has a thickness of 0.3nm. The thickness of the phonon layer controls the effective oscillator strength of the phonon mode. A thin layer is also analogous to having an adsorbed layer of molecules on the graphene nanoribbons, a setup commonly used in molecular sensing.

% Strong and weak coupling regimes
\par To determine the weak and strong coupling behavior of the graphene plasmon-polar dielectric phonon, we vary the distance, $s$, between the plasmon and phonon layers. The graphene plasmon frequency is tuned to match the SO phonon frequency via doping or electrostatic gating. The absorption spectra for strong and weak coupling in \cref{fig:schematic}(b) and (c) show lineshapes that are qualitatively different. In the strong coupling case, the absorption spectrum shows two distinct peaks (red arrows) that are well separated in frequency. The splitting in frequency is due to hybridization between the plasmon and phonon modes. The region of near-zero absorption between the modes naturally arises since there are no excitations in this spectral region. On the other hand, in the weak coupling case, we see a sharp transparency window in a broader absorption peak. This behavior is analogous to the electromagnetically induced transparency (EIT)\cite{Fleischhauer2005}, but in this case is produced by a destructive interference between the plasmon and phonon modes\cite{Yan2014, Low2014a}. While the absorption spectrum for both strong and weak coupling has a transparency region at $\omega_{SO}$, the corresponding field patterns shown in \cref{fig:schematic}(d) are clearly dissimilar. The strong coupling case shows negligible field excitation when compared to the weak coupling case in which the phonon and plasmon are both strongly excited. This observation is consistent since the transparency in the weak coupling case arises not from an absence of modes but rather from an interaction between the modes resulting in destructive interference.

\begin{figure}[h]
    \centering
    \includegraphics[width=0.5\textwidth, height=0.648\textwidth]{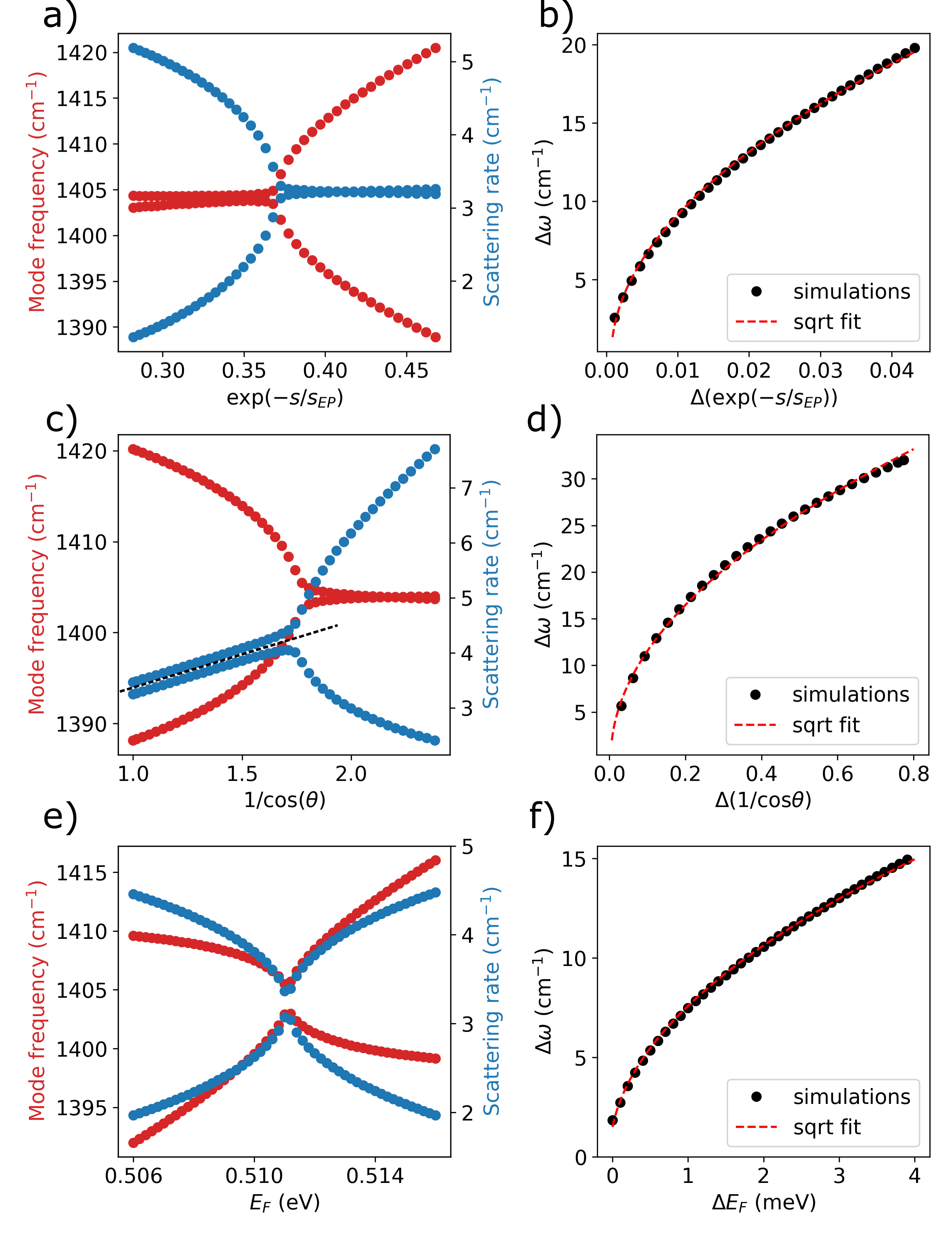}
    \caption{The eigenmode dispersion and mode frequency splitting as a function of different system parameters are plotted. The parameters used to modulate the system are spacer thickness, $s$ (a-b), incident angle of radiation, $\theta$ (c-d), and the graphene Fermi energy, $E_F$ (e-f).}
    \label{fig:eigenvalues_splitting}
\end{figure}

% Need for harmonic inversion analysis
\par Indeed, there has been considerable discussion on distinguishing between strong and weak light-matter coupling through its extinction/absorption spectrum\cite{Anisimov2011, Peng2014a, Hao2018, He2015}. A precise classification can be facilitated by using the harmonic inversion analysis method\cite{Fuchs2014}, which spectrally decompose the absorption spectrum into a superposition of complex Lorentzian resonances, with the form
\begin{equation}
    A(\omega) = \mbox{Im}\left(\sum_{i}\frac{d_{i}}{\omega-\omega_{i}-i\gamma_{i}}\right)
\end{equation}
where $d_{i}$ determines the lineshape of the decomposed resonance while $\omega_{i}$ and $\gamma_{i}$ are the resonance frequency and scattering rates respectivey. Results of the harmonic inversion analysis applied to absorption spectra are shown as dashed lines in \cref{fig:schematic} (b),(c). In the weak coupling limit the decomposed mode frequencies are degenerate with contrasting linewidths, as in EIT. For the strong coupling limit, mode frequencies are well separated, indicating hybridized modes. Discussion on the efficacy of the harmonic inversion method to decompose the absorption spectrum is given in the Supplemental Information (SI).

% Conditions for observing the exceptional point
As shown in \cref{fig:eigensurface}, the exceptional point can be observed under the variation of $\kappa$ (or $\alpha$) and $\omega_{pl}$. Since the EP is observed when $\omega_{pl}=\omega_{SO}$ and $\kappa=\gamma$, it is equally possible to encounter the EP under some parametric cut of $\gamma$ and $\omega_{pl}$. In the following, we show through analysis of the absorption spectrum that an EP may be observed under variation of the spacer thickness ($s$), incident angle of light ($\theta$), and the graphene Fermi level ($E_F$).

% Coupling strength modulation
\par To observe the transition from strong to weak coupling, we fix $E_{F}$ such that $\omega_{pl}(E_F)=\omega_{SO}$ and vary the thickness, $s$, between graphene nanoribbons and the polar dielectric, effectively tuning the coupling strength. Since the coupling strength is determined by the overlap of the plasmon and phonon modes we expect the coupling strength to exponentially decay with $s$ (i.e. $\kappa\propto\exp(-s)$). Other factors, such as $E_F$, which determine the confinement of the plasmon or phonon modes will also affect the coupling strength. By applying the harmonic inversion analysis to the absorption spectra calculated for a range of spacer thickness, $s$, we obtained \cref{fig:eigenvalues_splitting}(a). This dispersion closely matches the red path on the eigenvalue surface in \cref{fig:eigensurface} and may be identified as a PT-symmetry breaking transition. The EP can be identified as the boundary between strong and weak coupling regimes. \cref{fig:eigenvalues_splitting}(b) shows the square-root dependence of the mode frequencies splitting ($\Delta\omega$) on the strength of the perturbation $\Delta(\exp(-s/s^{EP})) = |\exp(-s/s^{EP})-\exp(-1)|$, which is a clear defining characteristic of the exceptional point.

% Angle dependence
Next, we show that tuning the angle of incidence ($\theta$, see \cref{fig:eigenvalues_splitting}(b)) provides an additional route to observing the exceptional point. By comparing expressions derived from the coupled mode equations in \cref{eq:coupled_mode_equations} and from conventional electromagnetic scattering calculations (see SI), we find that the loss of the plasmon is given by
\begin{equation}\label{eq:scattering_rate}
  	\gamma_{pl,i}=\frac{\gamma_{gr}}{2},\quad \gamma_{pl,e}=\frac{\mu_0cfD}{8\pi\cos\theta}
\end{equation}
where $\gamma_{pl,i}$ defines the intrinsic loss of the plasmon resonance, $\gamma_{pl,e}$ describes the radiative loss of the plasmon through coupling into free space, and $\gamma_{gr}$ is the plasmon scattering rate of graphene. Physically, the dependence on angle of $\gamma_{pl,e}$ accounts for the fact that the out coupling to free space depends on the angle that the plane wave momentum makes with the graphene ribbon surface\cite{Verslegers2010}. Having $\gamma_{pl,e}\propto (\cos\theta)^{-1}$ is required for isotropic out-coupling of the plasmon resonance to free space. The mode dispersion from simulations sweeping over a range of angles while fixing $\omega_{pl}=\omega_{SO}$ and $s=s^{EP}$ are shown in \cref{fig:eigenvalues_splitting}(c). A PT-symmetry breaking transition is shown as a function of $(\cos\theta)^{-1}$. The positive linear slope (dashed line in \cref{fig:eigenvalues_splitting}(c)) that we see in the scattering rate comes from the fact that as the angle is tuned, the average loss of the system, $\bar{\gamma}=(\gamma_{pl}+\gamma_{SO})/2$ increases linearly. This behavior is expected from \cref{eq:eigenvalues} and reassures us that we are indeed tuning the radiative loss of the plasmon.

% Fermi level modulation
\par We may also probe the EP by tuning the graphene Fermi level. If the plasmon frequency is continuously tuned such that it passes through $\omega_{SO}$, the coupled mode dispersion will exhibit level repulsion(level crossing) for $\kappa>\kappa^{EP}$($\kappa<\kappa^{EP}$). In \cref{fig:eigenvalues_splitting}(e), we show that by fixing $s=s^{EP}$ and varying the Fermi level, the EP is observed. Note that the shape of the dispersion is different from what is obtained in \cref{fig:eigenvalues_splitting}(a),(c). This is because by changing $\omega_{pl}$ we are breaking the PT-symmetry condition (except for when $\omega_{pl}=\omega_{SO}$). Also, tuning $E_{F}$ at fixed $s$ does not guarantee that the coupling strength is constant since $E_{F}$ also affects the plasmon confinement. Nevertheless, we are still able to find a path similar to the white line in \cref{fig:eigensurface} that passes through the EP. \cref{fig:eigenvalues_splitting}(f) verifies the square-root splitting for perturbations near the exceptional point. From these results we expect that electrostatic tuning of the graphene nanoribbons combined with exceptional point physics will provide a convenient route to realizing exceptional point sensors or modulators\cite{Wiersig2020a, Park2020}.

% Application of the plasmon-phonon coupled system to sensors
\section{Detection of the EP and use for optical sensing}
\begin{figure}
    \centering
    \includegraphics[width=0.5\textwidth, height=0.3846\textwidth]{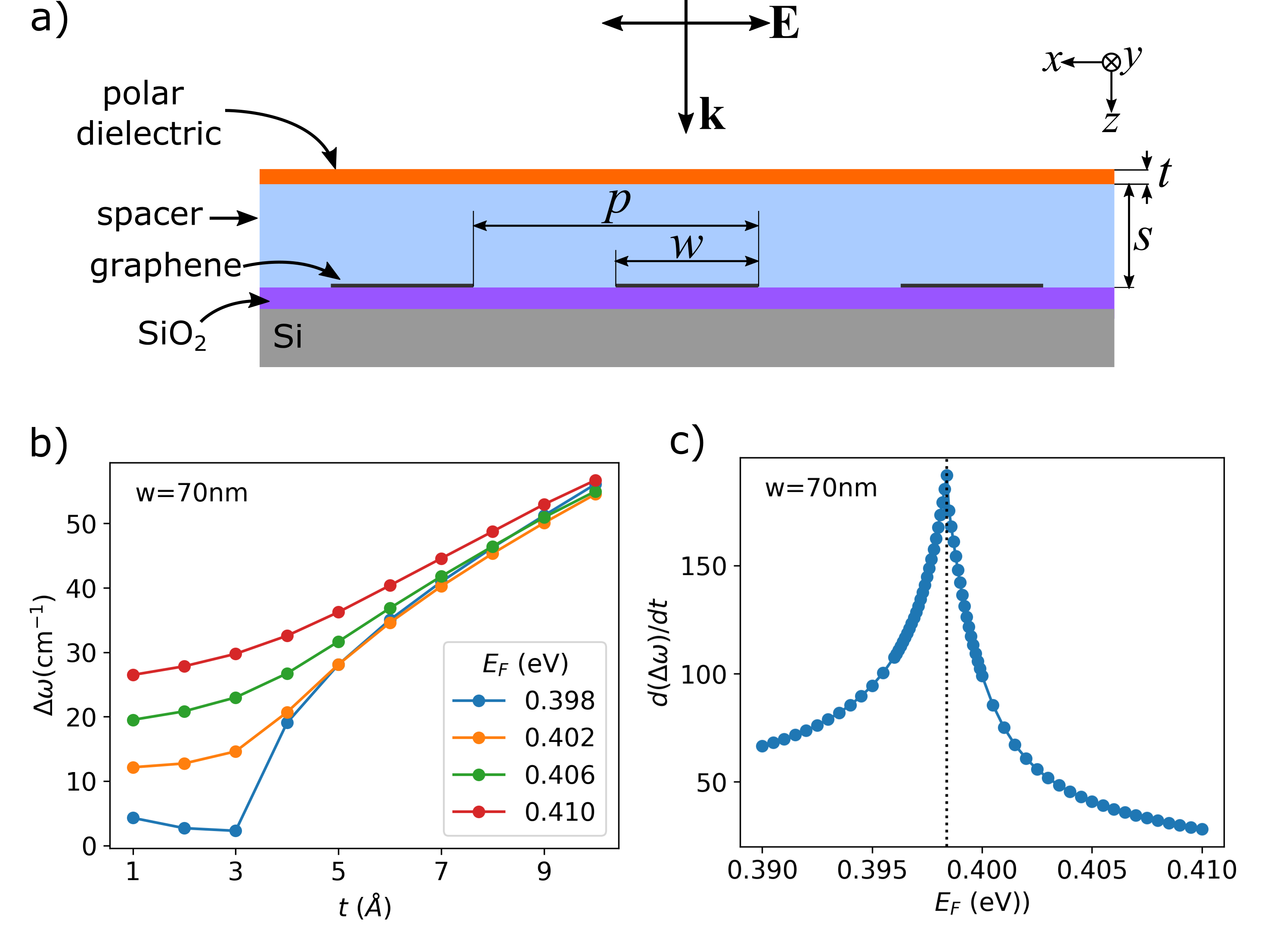}
    \caption{(a) Schematic for proposed thin film sensor at exceptional point. (b) Mode splitting as a function of thin film thickness. (c) Figure-of-merit as a function of $E_F$. Dashed line shows $E_F=0.398$eV$=E_F^{EP}$.}
    \label{fig:thickness_sensor}
\end{figure}
\par We propose a setup where the EP could be observed experimentally. At the EP, spectral splitting varies as the square root of perturbations\cite{Wiersig2020a}. Hence, we expect the rate of change of the splitting to diverge at the EP. We consider varying thickness in the phonon active layer. Coupling strength of the vibrational mode of the deposited layer with the graphene plasmon will depend on the thickness $t$. Thus when the system is tuned to the EP, the mode splitting will respond sensitively to small changes in the thickness. The proposed setup is shown in \cref{fig:thickness_sensor}(a). As a spacer we use an insulator with $n=1.36$ (e.g. CaF$_2$\cite{Malitson1963}) and a thin film layer with a phonon mode at $1400$cm$^{-1}$ is deposited on top. The spacer layer thickness is fixed to 27nm. We have also changed the ribbon width to 70nm such that the plasmon resonance is $\sim 1400$cm$^{-1}$ for $E_F\sim0.4$eV\cite{Rodrigo2015}. The structure is placed on a SiO$_2$/Si substrate to allow electrostatic gating via a back gate.

The mode splitting ($\Delta\omega$, see \cref{fig:thickness_sensor}(b)) shows a sharp change as a function of $t$ when $E_F=0.398$eV, while for higher $E_F$ the change is more gradual. The square-root splitting implies that for this system we have $E_F^{EP}\approx0.398$eV. To quantitatively gauge the sensitivity of our system, we define a figure-of-merit as $\partial(\Delta\omega)/\partial t|_{E_F}$ that measures the change in splitting with respect to a change in cover layer thickness at a fixed $E_F$. Indeed, we find that $\partial(\Delta\omega)/\partial t|_{E_F}$ is maximized when $E_F=E_F^{EP}$, as shown in \cref{fig:thickness_sensor}(c). Hence our setup utilizes the exceptional point to yield an enhanced sensitivity to the deposited dielectric layer. Similar setups can be used to sense changes produced by e.g. adsorption of molecules on the bare graphene nanoribbons or graphene modified with appropriate receptors.

\section{Conclusion}
We have demonstrated that tuning of the coupling or scattering rate of a plasmon-phonon system can lead to a PT-symmetry breaking transition with an EP observed at the boundary between weak and strong coupling. The coupling was controlled by the physical separation between the plasmon and polar dielectric layer while the scattering rate (i.e. its coupling to electromagnetic radiation) of the plasmon was shown to effectively be controlled by the incident angle of radiation. Importantly, when the coupling is set to a critical value, the characteristics of the exceptional point are also observable through electrostatic modulation of the graphene nanoribbons. Showing that the PT-symmetric transition and the EP are observable through experimentally accessible parameters might offer opportunities for sensitive optical sensors.

\begin{acknowledgements}
S.H.P. and T.L. acknowledge funding support from NSF/DMREF under Grant Agreement No. 1921629. S.-H.O. and T.L. acknowledge funding support from NSF ECCS No. 1809723.	
\end{acknowledgements}

\FloatBarrier
\bibliography{references}
\bibliographystyle{unsrt}
\clearpage
\end{document}